\documentclass[acmsmall,screen]{acmart}
\usepackage{xspace}
\usepackage{caption} 
\usepackage{tabularx}

\usepackage{amsmath,amsfonts}
\usepackage{algorithm}
\usepackage{algpseudocode}
\usepackage{float}
\usepackage{graphicx}
\usepackage{textcomp}
\usepackage{xcolor}

\usepackage{tcolorbox}
\tcbuselibrary{breakable}
\usepackage{multirow}
\usepackage{amsthm}
\usepackage{pifont}

\usepackage{booktabs}
\usepackage{array}
\usepackage{tikz}
\usepackage{colortbl}
\usepackage{makecell}

\newcommand{\tool}{TRACE\xspace}
\newcommand{\benchmark}{TRACEBench\xspace} 

\newcommand{\zsd}[1] {{\color{black}{#1}}}

\newcommand{\quartcirc}{%
\begin{tikzpicture}[baseline=-0.35ex]
\draw[fill=black!0] (0,0) circle (0.8ex);
\draw[fill=black] (0,0) -- (0,0.8ex) arc (90:0:0.8ex) -- cycle;
\end{tikzpicture}%
}
\newcommand{\halfcirc}{%
\begin{tikzpicture}[baseline=-0.35ex]
\draw[fill=black!0] (0,0) circle (0.8ex);
\draw[fill=black] (90:0.8ex) arc (90:-90:0.8ex) -- cycle;
\end{tikzpicture}%
}
\newcommand{\fullcirc}{%
\begin{tikzpicture}[baseline=-0.35ex]
\draw[fill=black] (0,0) circle (0.8ex);
\end{tikzpicture}%
}

\AtBeginDocument{%
  }

\setcopyright{none}
\begin{document}

\title{A Temporal Reasoning Benchmarking Framework for LRMs via Difficulty-controlled and Dynamic Test Generation}

\author{Shide Zhou}
\affiliation{%
  \institution{Huazhong University of Science and Technology}
  \city{Wuhan}
  \country{China}
}
\email{shidez@hust.edu.cn}

\author{Kailong Wang}
\authornote{Corresponding author.}
\affiliation{%
  \institution{Huazhong University of Science and Technology}
  \city{Wuhan}
  \country{China}
}
\affiliation{%
  \institution{National University of Singapore}
  \city{Singapore}
  \country{Singapore}
}
\email{wangkl@hust.edu.cn}

\author{Ling Shi}
\affiliation{%
  \institution{Nanyang Technological University}
  \city{Singapore}
  \country{Singapore}
}
\email{ling.shi@ntu.edu.sg}

\author{Haoyu Wang}
\affiliation{%
  \institution{Huazhong University of Science and Technology}
  \city{Wuhan}
  \country{China}
}
\email{haoyuwang@hust.edu.cn}

\begin{abstract}
Defining the reasoning boundaries and ensuring the reliability of Large Reasoning Models (LRMs) remains a critical challenge. Current benchmarks primarily rely on static datasets susceptible to data contamination or synthetic tasks lacking fine-grained difficulty control. Furthermore, standard outcome-based evaluations often conceal reasoning flaws by neglecting the reasoning process. 

To address these limitations, we introduce \tool{}, a testing framework that models temporal reasoning as constraint satisfaction problems via Allen's Interval Algebra. This approach enables precise regulation of logical complexity and incorporates a Trace-Based Verification Oracle to validate reasoning faithfulness. Using this framework, we construct \benchmark{}, an extensive benchmark comprising 1,200 synthesized test instances across graded difficulty levels. We employ \tool{} to evaluate eight widely used LRMs on \benchmark{}. The results confirm a strong negative correlation between model performance and our difficulty metric ($\text{Pearson's } r \approx -0.96$), validating the effectiveness of our difficulty control mechanism. 
Moreover, our trace-based analysis exposes significant discrepancies between reasoning validity and final answers, revealing a high spurious guessing rate of approximately 28\% in mid-sized models.
In addition, we diagnose scale-dependent failure modes, ranging from Degenerative Loops in small models to Reasoning Explosion in advanced architectures. \tool{} thus provides a robust, automated platform for benchmarking the true temporal reasoning capabilities of LRMs.
\end{abstract}



\maketitle

\section{Introduction}
The evolution of Large Language Models (LLMs) has culminated in the emergence of Large Reasoning Models (LRMs), such as the DeepSeek-R1 family~\cite{DBLP:journals/corr/abs-2501-12948, DBLP:journals/corr/abs-2412-19437} and OpenAI's o-series~\cite{openai2025}, designed specifically for complex problem-solving. In this domain, temporal reasoning is a fundamental capability, demanding strict logical consistency rather than approximate retrieval. While Chain-of-Thought(CoT) strategies~\cite{DBLP:conf/iclr/0002WSLCNCZ23, DBLP:conf/nips/Wei0SBIXCLZ22, DBLP:conf/nips/KojimaGRMI22} have yielded significant performance gains, a fundamental question persists: Do these improvements reflect genuine deduction or merely sophisticated pattern matching? This uncertainty complicates reliability assessments~\cite{DBLP:conf/icse/Du0WWL0FS0L24}, underscoring the urgent need for a specialized benchmarking framework.

\begin{table}[t]
\centering
\caption{Comparison of Our Work with Existing Benchmarks.}
\label{tab:comparison}
\begin{minipage}{0.8\columnwidth}
    \centering
    \resizebox{\textwidth}{!}{
        \begin{tabular}{cccccc}
        \toprule
         & \textbf{TRAM} & \textbf{TimeBench} & \textbf{\makecell{Test of Time}} & \textbf{t-BEN} & \cellcolor{gray!20}\textbf{Our Work} \\
        \midrule
        \textbf{Data Type} & \halfcirc & \halfcirc & \fullcirc & \fullcirc & \cellcolor{gray!20}\fullcirc \\
        \textbf{Difficulty Control} & \quartcirc & \quartcirc & \halfcirc & \halfcirc & \cellcolor{gray!20}\fullcirc \\
        \textbf{Verification Method} & \halfcirc & \halfcirc & \halfcirc & \halfcirc & \cellcolor{gray!20}\fullcirc \\
        \bottomrule
        \end{tabular}
    }
    \footnotesize
    \setlength{\baselineskip}{1.1em}
    \raggedright 
    \textbf{Data Type:} \halfcirc~Static Corpus; \fullcirc~Dynamic Synthetic.\\
    \textbf{Difficulty Control:} \quartcirc~Taxonomy-based; \halfcirc~Coarse-grained; \fullcirc~Fine-grained.\\
    \textbf{Verification Method:} \halfcirc~Outcome-based; \fullcirc~Dual Verification (Trace + Answer).
\end{minipage}
\vspace{-0.5cm}
\end{table}

Current benchmarks in temporal reasoning exhibit significant structural limitations, as summarized in Table~\ref{tab:comparison}. 
First, the reliance on static dataset aggregation in suites like TRAM~\cite{DBLP:conf/acl/Wang024} and TimeBench~\cite{DBLP:conf/acl/ChuCCY00024} introduces severe data contamination risks, allowing models to exploit memorization rather than engaging in genuine deduction~\cite{DBLP:conf/iclr/OrenMCLH24}. 
Second, while synthetic frameworks such as Test of Time~\cite{DBLP:conf/iclr/FatemiKTMYPSHP25} and t-BEN~\cite{wang2025tben} mitigate these risks, they typically employ coarse-grained difficulty proxies. Lacking precise regulation of logical complexity, these methods struggle to pinpoint specific breakdown points in a model's reasoning capabilities~\cite{DBLP:conf/iclr/ZhuC0GY024}. 
Third, prevailing evaluation methodologies remain strictly outcome-centric. By prioritizing final answer accuracy over process validity, they fail to detect spurious guessing, leaving the faithfulness of the reasoning trace largely unexamined~\cite{DBLP:conf/iclr/LightmanKBEBLLS24}.

\textbf{Our Work.} To overcome these structural deficits, a robust framework must fulfill three core design objectives: ensuring dynamic data generation to prevent contamination, implementing fine-grained difficulty control~\cite{DBLP:conf/emnlp/SinhaSDPH19} for precise boundary detection, and incorporating dual verification to validate both reasoning traces and final outcomes. To this end, we introduce \tool{} (\textbf{T}emporal \textbf{R}easoning \textbf{A}utomated \textbf{C}ontrollable \textbf{E}valuator). This framework builds upon Allen's Interval Algebra~\cite{10.1145/182.358434, DBLP:conf/aaai/Bessiere96}, which encompasses all 13 fundamental temporal relations, to model reasoning tasks as Constraint Satisfaction Problems (CSPs)~\cite{DBLP:conf/time/IsliB96}. \tool{} operates through three primary modules corresponding to the design objectives. The \textbf{Difficulty-Aware Constraint Generator} constructs constraint graphs. It allows users to strictly control logical complexity by setting a target difficulty and then adjusting the number of events and the types of temporal relations in the graph. The \textbf{Task Constructor} translates these graphs into natural language contexts. It utilizes the explicit edges of the graph as known premises and selects the implicit, inferred edges as questions to ensure the task requires reasoning. Finally, the \textbf{Trace-Based Verifier} assesses the model's logic. Unlike traditional methods that compare outputs against a fixed reference, this module verifies each step of the generated reasoning trace against the algebraic closure implied by the ground-truth constraint network.

We utilize \tool{} to construct \benchmark{}, a graded benchmark comprising 1,200 synthesized test instances spanning six distinct difficulty levels. \zsd{To ensure a focused assessment of LRMs' intrinsic logical reasoning rather than their tool-calling abilities, we purposely restrict the use of external solvers, thereby measuring pure deductive capabilities.} Using \benchmark{}, we conduct a comprehensive evaluation of eight widely used LRMs, ranging from distilled variants to advanced proprietary models. The results confirm the precision of our difficulty modeling. We observe a strong negative correlation between the difficulty score and model accuracy, with Pearson's r approximately $-0.96$. This result validates that the framework effectively generates tasks with a controllable gradient of complexity. Notably, our trace-based analysis exposes significant discrepancies between reasoning validity and final answer correctness. We find that mid-sized models exhibit a high spurious guessing rate of approximately 28\%, where they frequently arrive at the correct final label despite relying on invalid reasoning steps. Conversely, smaller architectures suffer from Answer Misalignment, where valid logic leads to incorrect final labels. Furthermore, we diagnose scale-dependent structural failure modes under extreme complexity. Small and mid-sized models often fall into Degenerative Loops of repetitive generation, whereas advanced models mainly face Reasoning Explosion, with valid reasoning chains becoming too long and eventually exceeding the context window.

\textbf{Contributions.} In summary, this work makes the following contributions:
\begin{itemize}
    \item \textbf{A difficulty-controllable testing framework.} We introduce \tool{}, a novel framework that leverages graph-based generation and Allen's Interval Algebra to synthesize logically consistent temporal reasoning tasks with precisely tunable complexity, incorporating a trace-based oracle for faithfulness verification.
    \item \textbf{An extensive graded benchmark.} We construct \benchmark{}, containing 1,200 instances across six difficulty levels, designed to systematically probe the reasoning capabilities of LRMs.
    \item \textbf{Empirical insights and failure taxonomy.} Through extensive evaluation, we uncover a high spurious guessing rate in mid-sized models and identify scale-dependent failure modes, including Degenerative Loops in smaller models and Reasoning Explosion in advanced models.
\end{itemize}

\section{Background and Related Work}
\subsection{Large Reasoning Model}
\label{sec:ReasoningModel}
The recent shift from scaling training parameters to scaling test-time computation has enabled the emergence of LRMs~\cite{DBLP:journals/corr/abs-2407-21787, DBLP:journals/corr/abs-2408-03314}. Unlike standard models that generate immediate responses, LRMs utilize an extended inference phase to decompose complex problems into intermediate logical steps. While initial capabilities are elicited through prompt engineering methods like Chain-of-Thought~\cite{DBLP:conf/iclr/0002WSLCNCZ23, DBLP:conf/nips/Wei0SBIXCLZ22, DBLP:conf/nips/KojimaGRMI22, DBLP:conf/nips/YaoYZS00N23, DBLP:conf/iclr/FuPSCK23}, state-of-the-art models such as the DeepSeek-R1 family~\cite{DBLP:journals/corr/abs-2501-12948, DBLP:journals/corr/abs-2412-19437} internalize this deliberative process through specialized post-training optimization, including Reinforcement Learning and rejection sampling~\cite{DBLP:conf/nips/ZelikmanWMG22, DBLP:conf/iclr/LightmanKBEBLLS24}. As a result, these models produce an output structured into two clear parts: an internal thinking process that traces intermediate reasoning and a final conclusion representing the definitive outcome.

\subsection{Evolution of Reasoning Benchmarks}
\label{sec:LLMEvaluation}
The evaluation of reasoning in language models has evolved from general competency tests to specialized, logic-intensive benchmarks. Initial efforts focused on static datasets designed to probe mathematical and symbolic reasoning. Canonical benchmarks such as GSM8K~\cite{DBLP:journals/corr/abs-2110-14168} and MATH~\cite{DBLP:conf/nips/HendrycksBKABTS21} established the standard for evaluating multi-step mathematical derivation. Similarly, BIG-Bench Hard~\cite{DBLP:conf/acl/SuzgunSSGTCCLCZ23} and MMLU~\cite{DBLP:conf/iclr/HendrycksBBZMSS21} extended this scope to encompass broad cross-domain reasoning tasks, while specialized datasets like SCAN~\cite{DBLP:conf/icml/LakeB18} and CLUTRR~\cite{DBLP:conf/emnlp/SinhaSDPH19} focused on compositional and inductive logic, respectively.

To mitigate the inherent risks of contamination and coarse granularity in such static datasets, recent research has shifted towards synthetic and difficulty-controllable evaluation frameworks. Approaches such as PrOntoQA~\cite{DBLP:conf/iclr/Saparov023} and DyVal~\cite{DBLP:conf/iclr/ZhuC0GY024} utilize generated logic puzzles to prevent memorization. More recently, GRADE~\cite{DBLP:journals/corr/abs-2508-16994} advanced this direction in retrieval-augmented systems by modeling task difficulty via a 2D matrix of reasoning depth and semantic distance, validating that error rates strictly follow structural constraints.

In the specific domain of temporal reasoning, benchmarks have progressed from static aggregation to dynamic synthesis. Works such as TRAM~\cite{DBLP:conf/acl/Wang024} construct comprehensive evaluation suites by unifying multiple datasets covering aspects like frequency, duration, and event ordering. Similarly, TimeBench~\cite{DBLP:conf/acl/ChuCCY00024} introduces a hierarchical taxonomy, organizing tasks into symbolic, commonsense, and event-based reasoning categories to probe different dimensions of temporal logic. To address static limitations, recent efforts such as Test of Time~\cite{DBLP:conf/iclr/FatemiKTMYPSHP25} and t-BEN~\cite{wang2025tben} have advanced this field by leveraging synthetic generation mechanisms, creating dynamic datasets to mitigate the memorization risks inherent in static corpora. Despite these advancements, current methodologies largely remain outcome-centric or rely on coarse-grained difficulty proxies, leaving the critical need for fine-grained complexity control and trace-level faithfulness verification unaddressed.


\subsection{Allen Interval Algebra}
\label{subsubsec:preliminaries}

\begin{table}[htbp]
\centering
\caption{The 13 Basic Relations in Allen Interval Algebra.}
\label{tab:allen_relations}
\resizebox{0.8\columnwidth}{!}{%
\begin{tabular}{cccc}
\toprule
\textbf{Relation / Inverse Relation} & \textbf{Symbol (Rel/Inv)} & \textbf{Visualization} & \textbf{Endpoint Conditions} \\ \midrule

before / after & B / BI & 
\begin{tikzpicture}[baseline=0.5ex, xscale=0.35, yscale=0.35]
    \draw[fill=gray!25] (0,0.5) rectangle (1.2,1.0); \node[scale=0.7] at (0.6, 0.75) {X};
    \draw[fill=white, draw=black] (1.6,0) rectangle (2.8,0.5); \node[scale=0.7] at (2.2, 0.25) {Y};
    \draw[dotted] (1.2,0.75) -- (1.6,0.25);
\end{tikzpicture} & 
$X_{e} < Y_{s}$ \\

meets / met-by & M / MI & 
\begin{tikzpicture}[baseline=0.5ex, xscale=0.35, yscale=0.35]
    \draw[fill=gray!25] (0,0.5) rectangle (1.2,1.0); \node[scale=0.7] at (0.6, 0.75) {X};
    \draw[fill=white, draw=black] (1.2,0) rectangle (2.4,0.5); \node[scale=0.7] at (1.8, 0.25) {Y};
    \draw (1.2,0) -- (1.2,1.0);
\end{tikzpicture} & 
$X_{e} = Y_{s}$ \\

overlaps / overlapped-by & O / OI & 
\begin{tikzpicture}[baseline=0.5ex, xscale=0.35, yscale=0.35]
    \draw[fill=gray!25] (0,0.5) rectangle (1.5,1.0); \node[scale=0.7] at (0.75, 0.75) {X};
    \draw[fill=white, draw=black] (1.0,0) rectangle (2.5,0.5); \node[scale=0.7] at (1.75, 0.25) {Y};
    \draw[dotted] (1.0,0) -- (1.0,1.0); \draw[dotted] (1.5,0) -- (1.5,1.0);
\end{tikzpicture} & 
$X_{s} < Y_{s} < X_{e} < Y_{e}$ \\

starts / started-by & S / SI & 
\begin{tikzpicture}[baseline=0.5ex, xscale=0.35, yscale=0.35]
    \draw[fill=gray!25] (0,0.5) rectangle (1.2,1.0); \node[scale=0.7] at (0.6, 0.75) {X};
    \draw[fill=white, draw=black] (0,0) rectangle (2.0,0.5); \node[scale=0.7] at (1.0, 0.25) {Y};
    \draw (0,0) -- (0,1.0);
\end{tikzpicture} & 
$X_{s} = Y_{s}, X_{e} < Y_{e}$ \\

during / contains & D / DI & 
\begin{tikzpicture}[baseline=0.5ex, xscale=0.35, yscale=0.35]
    \draw[fill=gray!25] (0.8,0.5) rectangle (2.0,1.0); \node[scale=0.7] at (1.4, 0.75) {X};
    \draw[fill=white, draw=black] (0,0) rectangle (2.8,0.5); \node[scale=0.7] at (1.4, 0.25) {Y};
\end{tikzpicture} & 
$Y_{s} < X_{s}, X_{e} < Y_{e}$ \\

finishes / finished-by & F / FI & 
\begin{tikzpicture}[baseline=0.5ex, xscale=0.35, yscale=0.35]
    \draw[fill=gray!25] (1.2,0.5) rectangle (2.4,1.0); \node[scale=0.7] at (1.8, 0.75) {X};
    \draw[fill=white, draw=black] (0,0) rectangle (2.4,0.5); \node[scale=0.7] at (1.2, 0.25) {Y};
    \draw (2.4,0) -- (2.4,1.0);
\end{tikzpicture} & 
$X_{e} = Y_{e}, X_{s} > Y_{s}$ \\

equals & E & 
\begin{tikzpicture}[baseline=0.5ex, xscale=0.35, yscale=0.35]
    \draw[fill=white, draw=black] (0,0) rectangle (1.5,0.5); 
    \draw[fill=gray!25] (0,0.5) rectangle (1.5,1.0);
    \node[scale=0.7] at (0.75, 0.75) {X}; \node[scale=0.7] at (0.75, 0.25) {Y};
\end{tikzpicture} & 
$X_{s} = Y_{s}, X_{e} = Y_{e}$ \\ 

\bottomrule
\end{tabular}%
}
\vspace{-0.5cm}
\end{table}

\tool{} is built upon Allen's Interval Algebra~\cite{10.1145/182.358434}, which treats time intervals as primitive objects and provides a robust reasoning framework for temporal relations. An interval $X$ is defined by its start and end points, denoted as $[X_{s}, X_{e}]$ where $X_{s} < X_{e}$. The algebra defines a set of 13 mutually exclusive basic relations that capture all possible qualitative positions between two intervals. As shown in Table~\ref{tab:allen_relations}, these include six asymmetric relations (e.g., \textit{before}) with their corresponding inverses (e.g., \textit{after}), and one symmetric relation (\textit{equals}). A temporal constraint is formed by assigning one of these relations to a pair of intervals (e.g., $X$ \textit{overlaps} $Y$). Consequently, a reasoning task can be modeled as a network where each node corresponds to an event represented as a time interval, and each edge encodes a temporal constraint.

A fundamental property of this algebra lies in its ability to support deductive reasoning through transitivity. Intuitively, if event $X$ happens \textit{before} $Y$, and $Y$ happens \textit{before} $Z$, logic dictates that $X$ must be \textit{before} $Z$. Formally, this is determined by the composition of relations: Given two constraints $\smash{X \xrightarrow{r_1} Y}$ and $\smash{Y \xrightarrow{r_2} Z}$, where $r_1, r_2 \in \mathcal{R}_{Allen}$ denote the specific relations, the valid relationship between $X$ and $Z$ is constrained to the composition set $r_1 \circ r_2 \subseteq \mathcal{R}_{Allen}$.

To serve as a valid reasoning task, a constraint graph must strictly satisfy \textit{Path Consistency}. We enforce path consistency as a generation-time constraint: for every triple of nodes $(i, j, k)$, the direct constraint between $i$ and $k$ does not conflict with the transitive constraints inferred through $j$. We apply this property to rule out impossible scenarios (e.g., a cycle where $A$ is before $B$ and $B$ is before $A$), ensuring the resulting instances are logically contradiction-free and admit a consistent temporal interpretation.
\section{Methodology}
\label{sec:methodology}
In this section, we present the design and implementation of \tool{}, a difficulty-controllable testing framework specifically tailored for evaluating LRMs in complex temporal reasoning scenarios. \tool{} enables the dynamic generation of reasoning tasks with tunable complexity, coupled with a rigorous verification mechanism to assess the faithfulness of the reasoning process. We first provide a high-level overview of the system architecture. Subsequently, we elaborate on the three core modules: Difficulty-Aware Constraint Graph Generation, Temporal Reasoning Task Construction, and the Trace-Based Verification Oracle.

\subsection{Overview of \tool{}}
\label{subsec:overview}
\begin{figure}[htbp]
    \vspace{-20pt}
    \centering
    \includegraphics[width=\textwidth]{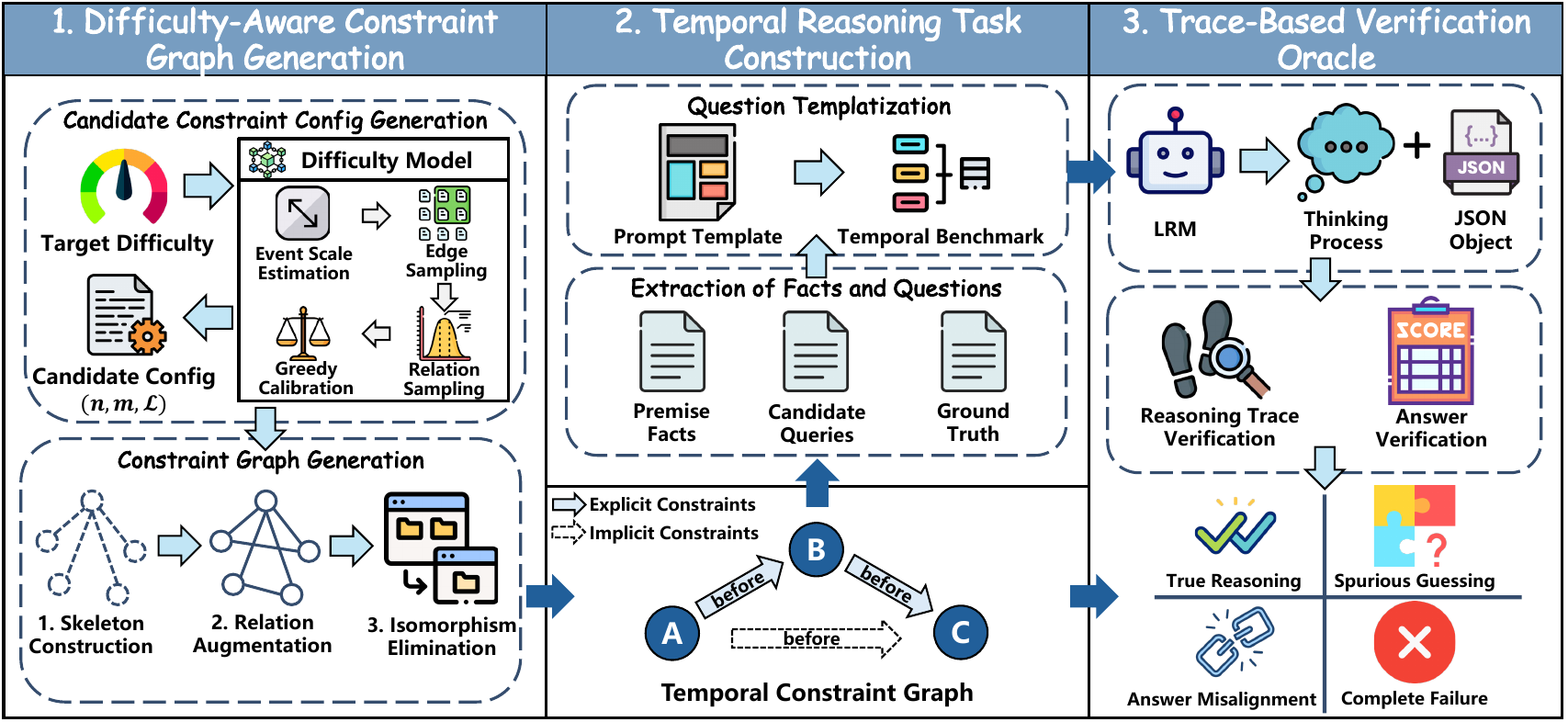}
    \caption{\zsd{The Workflow of \tool{}.}}
    \label{fig:overview}
    \vspace{-20pt}
\end{figure}

The overall architecture of \tool{} is illustrated in Figure~\ref{fig:overview}. The workflow begins with a difficulty configuration, where the user sets a target difficulty level. Based on this configuration, the Difficulty-Aware Constraint Generator~(Section~\ref{subsec:generation}) constructs a series of mathematically consistent constraint graphs, which serve as the logical backbone of the test cases. These abstract graphs are then processed by the Task Constructor~(Section~\ref{subsec:construction}), which translates the constraints into a natural language context. Specifically, the system formulates questions based on implicit relations, which are logical consequences not explicitly stated in the context. This design ensures that the tasks strictly require deductive reasoning.

In the final stage, the generated tasks are executed by the target LRM. \zsd{During this phase, the model operates independently without access to external tools or solvers, ensuring the evaluation isolates pure internal reasoning.} Through structured prompting, the model is required to output both the formal reasoning process and the final answer. \tool{} then employs a Trace-Based Verifier~(Section~\ref{subsec:oracle}) to extract the reasoning traces from the model's responses and uses a constraint solver to validate the correctness of each reasoning step. This mechanism enables \tool{} to distinguish genuine reasoning from spurious guesses, comprehensively assessing reasoning faithfulness.

\subsection{Difficulty-Aware Constraint Graph Generation}
\label{subsec:generation}

\subsubsection{Difficulty Modeling}
\label{subsubsec:difficulty}
To systematically quantify the difficulty of a temporal reasoning task, we propose a model based on two primary dimensions: \emph{event-set scale} and \emph{constraint complexity}.

\textbf{Problem Formalization.}
We model a temporal reasoning task as a constraint graph $\mathcal{G} = (V, E)$, where $V = \{v_1, v_2, \dots, v_n\}$ denotes the set of $n$ unique events and $E$ denotes the set of temporal constraints given as premises. Each constraint $(v_i, v_j) \in E$ is labeled with a relation type $r \in \mathcal{R}_{Allen}$ (e.g., \textit{B for before}, \textit{O for overlaps}). Note that $E$ contains only the explicit constraints provided as context, excluding any implicit relations to be inferred.

\textbf{Difficulty Definition.}
We posit that task difficulty is jointly determined by event-set scale and constraint complexity. Specifically, we define the difficulty score $\mathcal{D}$ for a graph $\mathcal{G}$ as:

\begin{equation}
    \mathcal{D}(\mathcal{G}) = |V|^{\alpha} \cdot \underbrace{\left( \frac{1}{|E|} \sum_{(v_i, v_j) \in E} w(r_{ij}) \right)}_{\overline{w}}
    \label{eq:difficulty_model}
\end{equation}

where $|V|$ is the number of events (event-set scale), $\alpha$ controls how scale amplifies difficulty, and $w(r)$ maps each relation type to a scalar complexity score.

\textbf{Design Rationale.}
This formulation captures two orthogonal components of difficulty: \textbf{\ding{182} Event-set scale ($|V|^{\alpha}$):} As the number of events increases, the number of possible event permutations grows super-linearly. The exponent $\alpha$ models this expansion, reflecting how the increase in context length and reasoning depth impacts the difficulty. \textbf{\ding{183} Constraint complexity ($\overline{w}$):} Not all relations are equally hard to process. The term $\overline{w}$ represents the average inference cost per edge. For instance, determining a precise ordering under a complex constraint (e.g., \textit{overlaps}) requires handling more uncertainty than a simple constraint (e.g., \textit{equals}). This term distinguishes graphs with the same number of events but different average constraint complexity.

\textbf{Complexity Analysis and Model Calibration.}
To instantiate the difficulty model, we assign a weight $w(r)$ to each Allen relation and calibrate the scaling exponent $\alpha$. Our weights reflect the cognitive load required to maintain path consistency, determined by the degrees of freedom and boundary precision. Relations that reduce the number of independent variables are treated as easier, while those requiring the simultaneous satisfaction of multiple coupled inequalities without explicit anchors are considered harder.

\textit{1) Relation Complexity Weights.}
We group the 13 Allen relations into four tiers based on their impact on the reasoning state space; for each inverse pair, we use the same weight, i.e., $w(r)=w(r^{-1})$:
\begin{itemize}
    \item \textbf{Coincidence Constraints ($w=0.8$):}
    The equals ($E$) relation enforces that two events share identical start and end points ($Start_A = Start_B \land End_A = End_B$). Although this involves two conditions, logically, it collapses two distinct events into a single temporal entity. This effective dimensionality reduction simplifies the constraint graph, lowering the memory and inference burden required to track the timeline.

    \item \textbf{Precedence Constraints ($w=1.0$):}
    Relations \textit{before} ($B$) and \textit{after} ($BI$) express a basic precedence condition (e.g., $End_A < Start_B$). Given its high degrees of freedom, we treat this ubiquitous “A before B” pattern as the baseline unit of difficulty.

    \item \textbf{No-gap Adjacency Constraints ($w=1.1$):}
    Relations \textit{meets} ($M$) and \textit{met-by} ($MI$) strengthen precedence by requiring exact boundary alignment (e.g., $End_A = Start_B$). We assign a slightly higher weight to reflect the additional precision needed to satisfy and verify this zero-gap condition.

    \item \textbf{Endpoint-interleaving Constraints ($w\in\{1.5,2.0\}$):} Relations such as \textit{starts} ($S$) and \textit{finishes} ($F$) include an endpoint equality (e.g., $Start_A = Start_B$ or $End_A = End_B$) together with an inequality, which partially anchors the configuration; we assign them $w=1.5$. In contrast, relations such as \textit{overlaps} ($O$) and \textit{during} ($D$) require satisfying multiple coupled inequalities simultaneously (e.g., \textit{overlaps}: $Start_A < Start_B < End_A < End_B$), which is typically harder to deduce; we assign them $w=2.0$.
\end{itemize}


Although extreme cases (e.g., all \textit{equals}) may slightly distort the difficulty score, our strategy of diverse constraint sampling helps the score better reflect the actual reasoning complexity in practice.

\textit{2) Calibrating the Scale Exponent.}
To set $\alpha$ in an interpretable way, we anchor the metric to a \textbf{standard reference task}: the smallest transitive inference problem in which three events are given with two explicit relations, and the model must infer the third relation. We define the difficulty of this atomic task as $\mathcal{D}_{ref}=10$. Assuming a uniform distribution over relation types, the mean relation weight is $\overline{w} \approx 1.46$. Substituting into our formula yields $10 = 3^{\alpha} \cdot 1.46$, and solving gives $\alpha \approx 1.75$. This calibration grounds the metric in a solvable baseline case and ensures difficulty grows smoothly with problem scale.

\subsubsection{Candidate Constraint Configuration Generation}
\label{subsubsec:candidate_config}

\begin{algorithm}[htbp]
\caption{\zsd{Candidate Constraint Configuration Generation}}
\label{alg:candidate_config}
\zsd{
\begin{algorithmic}[1]
\Require Target difficulty $\mathcal{D}_{\mathrm{tar}}$, relation set $\mathcal{R}_{Allen}$ with weights $w(\cdot)$, scale exponent $\alpha$, slack $\Delta$,  density $\rho$, softmax width $\sigma$, tolerance $\tau$, max adjust steps $T$.
\Ensure Candidate configuration $(n,m,\mathcal{L})$.

\State \textbf{Phase 1: Event Scale Estimation \& Edge Sampling}
\State $\overline{w}_{\mathrm{prior}} \gets \frac{1}{|\mathcal{R}_{Allen}|}\sum_{r\in \mathcal{R}_{Allen}} w(r)$; 
\quad $n \gets \max\!\left(3,\; \mathrm{round}\!\left((\mathcal{D}_{\mathrm{tar}}/\overline{w}_{\mathrm{prior}})^{1/\alpha}\right)\right)$

\State $N_{\mathrm{pair}} \gets \binom{n}{2}$; \quad $m_{\mathrm{exp}} \gets \mathrm{round}(\rho\cdot N_{\mathrm{pair}})$
\State Compute bounds $[m_{\mathrm{lo}}, m_{\mathrm{hi}}]$ around $m_{\mathrm{exp}}$
\State Sample $m \sim \mathrm{Unif}(\{m_{\mathrm{lo}},\ldots,m_{\mathrm{hi}}\})$

\State \textbf{Phase 2: Relation Sampling}
\State $\overline{w}_{\mathrm{tar}} \gets \mathcal{D}_{\mathrm{tar}}/n^{\alpha}$; \quad Compute $p(r) \propto \exp(-(w(r)-\overline{w}_{\mathrm{tar}})^2 / 2\sigma^2)$
\State Sample multiset $\mathcal{L}$ of size $m$ from distributions $p(\cdot)$

\State \textbf{Phase 3: Greedy Calibration}
\State $\overline{w}_{\mathrm{ach}} \gets \frac{1}{m}\sum_{r\in\mathcal{L}} w(r)$; \quad$\varepsilon \gets \frac{|\overline{w}_{\mathrm{ach}}-\overline{w}_{\mathrm{tar}}|}{\overline{w}_{\mathrm{tar}}}$; \quad $t \gets 0$

\While{$\varepsilon > \tau$ \textbf{and} $t < T$}
    \State Swap one extreme-weight relation in $\mathcal{L}$ to move $\overline{w}_{\mathrm{ach}}$ towards $\overline{w}_{\mathrm{tar}}$
    \State Update $\overline{w}_{\mathrm{ach}}$ and $\varepsilon$; \quad $t \gets t + 1$
\EndWhile
\State \Return $(n,m,\mathcal{L})$
\end{algorithmic}
}
\end{algorithm}

Based on the difficulty model proposed in Section~\ref{subsubsec:difficulty}, \tool{} first converts the target difficulty $\mathcal{D}_{\mathrm{tar}}$ into a candidate constraint configuration that is expected to yield an achieved difficulty $\mathcal{D}_{\mathrm{ach}}$ close to $\mathcal{D}_{\mathrm{tar}}$. Algorithm~\ref{alg:candidate_config} summarizes the complete procedure. Concretely, a configuration is represented as a tuple $(n,m,\mathcal{L})$, where $n$ is the number of events, $m$ is the number of explicit constraints to be presented as premises, and $\mathcal{L}$ is a multiset of $m$ Allen relations.

The generation starts by inferring a reasonable event-set size. Let $\overline{w}_{\mathrm{prior}}$ denote the average weight over the allowed relation set,
\begin{equation}
\overline{w}_{\mathrm{prior}}=\frac{1}{|\mathcal{R}_{Allen}|}\sum_{r\in \mathcal{R}_{Allen}} w(r).
\end{equation}
Following Eq.~\eqref{eq:difficulty_model}, we invert the relationship between scale and difficulty to estimate the event count:
\begin{equation}
n = \max\left(3,\mathrm{round}\left(\left(\frac{\mathcal{D}_{\mathrm{tar}}}{\overline{w}_{\mathrm{prior}}}\right)^{\frac{1}{\alpha}}\right)\right)
\end{equation}
The lower bound of three guarantees the smallest setting in which transitive inference is meaningful.

Given $n$, \tool{} determines the number of explicit constraints $m$ by controlling the edge density. The number of possible unordered event pairs is $N_{\mathrm{pair}}=\binom{n}{2}$. When $m$ is too large (close to $N_{\mathrm{pair}}$), the context becomes nearly complete and leaves little room for deductive inference; when $m$ is too small, the graph tends to be under-constrained or disconnected, which makes deterministic reasoning unreliable. We therefore use a density parameter $\rho\in(0,1)$ to set an expected edge count $m_{\mathrm{exp}}=\mathrm{round}(\rho\cdot N_{\mathrm{pair}})$ and sample $m$ from a bounded interval around it:
\begin{align}
m_{\mathrm{lo}} &= \max(n-1,\; m_{\mathrm{exp}}-\Delta), \\
m_{\mathrm{hi}} &= \min(N_{\mathrm{pair}}-1,\; m_{\mathrm{exp}}+\Delta), \\
m &\sim \mathrm{Unif}\big(\{m_{\mathrm{lo}},\ldots,m_{\mathrm{hi}}\}\big),
\end{align}
where $\Delta$ is a small slack constant. The lower bound $n-1$ encourages connectivity, while the upper bound $N_{\mathrm{pair}}-1$ avoids producing a fully saturated graph.

After fixing $(n,m)$, \tool{} computes the target mean relation weight implied by Eq.~\eqref{eq:difficulty_model}:
\begin{equation}
\overline{w}_{\mathrm{tar}} \;=\; \frac{\mathcal{D}_{\mathrm{tar}}}{n^{\alpha}}.
\end{equation}
Intuitively, $\overline{w}_{\mathrm{tar}}$ specifies the average constraint complexity required at scale $n$ to match the target difficulty. To bias relation sampling toward this target, \tool{} assigns each relation $r$ a probability based on the distance between $w(r)$ and $\overline{w}_{\mathrm{tar}}$. Specifically, we use a Gaussian-shaped softmax:
\begin{equation}
p(r)\propto \exp\!\left(-\frac{(w(r)-\overline{w}_{\mathrm{tar}})^2}{2\sigma^2}\right)
\end{equation}
so that relations with weights closer to $\overline{w}_{\mathrm{tar}}$ are more likely to be selected. We then draw $m$ relations from this distribution to obtain an initial multiset $\mathcal{L}$.

Because finite sampling can deviate from the intended mean, \tool{} measures the relative error between the achieved mean weight $\overline{w}_{\mathrm{ach}}$ of $\mathcal{L}$ and $\overline{w}_{\mathrm{tar}}$:
\begin{equation}
\varepsilon \;=\; \frac{|\overline{w}_{\mathrm{ach}}-\overline{w}_{\mathrm{tar}}|}{\overline{w}_{\mathrm{tar}}}.
\end{equation}
If $\varepsilon$ exceeds a preset tolerance $\tau$, we apply a greedy calibration to adjust the multiset composition while keeping $m$ fixed. When $\overline{w}_{\mathrm{ach}}>\overline{w}_{\mathrm{tar}}$, we repeatedly replace one occurrence of a currently highest-weight relation with the minimum-weight relation; when $\overline{w}_{\mathrm{ach}}<\overline{w}_{\mathrm{tar}}$, we symmetrically replace one minimum-weight relation with a maximum-weight relation. This bounded adjustment monotonically moves $\overline{w}_{\mathrm{ach}}$ toward $\overline{w}_{\mathrm{tar}}$ and yields the final relation multiset $\mathcal{L}$.

Finally, \tool{} computes the achieved difficulty
$\mathcal{D}_{\mathrm{ach}} \;=\; n^{\alpha}\cdot \overline{w}_{\mathrm{ach}}$
and outputs the candidate configuration $(n,m,\mathcal{L})$, which is instantiated into a path-consistent constraint graph in the next stage.

\subsubsection{Constraint Graph Generation} 
\label{subsubsec:graph_generation}

\begin{algorithm}[htbp]
\caption{\zsd{Path-Consistent Constraint Graph Construction}}
\label{alg:graph_generation}
\zsd{
\begin{algorithmic}[1]
\Require Candidate configuration $(n,m,\mathcal{L})$, History signatures $\Sigma$.
\Ensure Path-consistent constraint graph $\mathcal{G}$, Canonical signature $\mathcal{S}$.

\State \textbf{Phase 1: Skeleton Construction}
\State Initialize $\mathcal{G} \gets (V, E_{\mathrm{tree}})$ via random Prufer sequence spanning tree
\State Partition $\mathcal{L}$ into $\mathcal{L}_{\mathrm{skel}}$ (size $n-1$) and $\mathcal{L}_{\mathrm{rem}}$
\State Map $\mathcal{L}_{\mathrm{skel}}$ to $E_{\mathrm{tree}}$; \quad Compute degrees $\deg(v)$ for all $v \in V$

\State \textbf{Phase 2: Relation Augmentation}
\State $P_{\mathrm{free}} \gets$ Unconnected pairs sorted by ascending cost $k(u,v) = \deg(u) + \deg(v)$
\For{$r \in \mathcal{L}_{\mathrm{rem}}$}
    \State Find first $(u,v) \in P_{\mathrm{free}}$ satisfying $\mathrm{IsConsistent}(\mathcal{G} \cup \{(u,v,r)\})$
    \If{no valid pair found} \Return \textbf{Failure} \EndIf
    \State Add $(u,v,r)$ to $\mathcal{G}$; \quad Update degrees; \quad Remove $(u,v)$ from $P_{\mathrm{free}}$
\EndFor

\State \textbf{Phase 3: Isomorphism Elimination}
\If{\textbf{not} $\mathrm{IsConsistent}(\mathcal{G})$} \Return \textbf{Failure} \EndIf
\State $\mathcal{S} \gets \mathrm{CanonicalHash}(\mathcal{G})$ 
\If{$\mathcal{S} \in \Sigma$} \Return \textbf{Failure} \EndIf
\State \Return $(\mathcal{G}, \mathcal{S})$
\end{algorithmic}%
}
\end{algorithm}

Given the candidate constraint configuration $\mathcal{C} = (n, m, \mathcal{L})$ derived in Section~\ref{subsubsec:candidate_config}, our objective is to instantiate these parameters into a concrete, logically consistent temporal constraint graph $\mathcal{G} = (V, E)$. This process is non-trivial; a naive stochastic assignment of relations to arbitrary event pairs frequently induces logical contradictions (e.g., a cycle $\smash{A \xrightarrow{\text{before}} B \xrightarrow{\text{before}} A}$), rendering the graph mathematically invalid. To address this, we propose a constructive generation procedure, formalized in Algorithm~\ref{alg:graph_generation}, which builds the network incrementally to guarantee satisfiability.

The procedure initiates with \textbf{Skeleton Construction}. Our primary goal is to establish a connected backbone for the event set $V=\{v_1, \dots, v_n\}$ without introducing structural conflicts. To ensure that every event is reachable and integrated into the reasoning chain, we generate a random spanning tree $E_{tree}$ consisting of $n-1$ edges based on a Prufer sequence.\footnote{The Prufer sequence provides a bijection between the set of labeled trees on $n$ vertices and sequences of length $n-2$, thereby ensuring an unbiased sampling of tree topologies.}

Subsequently, we align the relation set with this topology. We partition the candidate relation multiset $\mathcal{L}$ into a backbone set $\mathcal{L}_{skel}$ and a remainder set $\mathcal{L}_{rem}$:
\begin{equation}
\mathcal{L} = \mathcal{L}_{skel} \uplus \mathcal{L}_{rem}, \quad \text{where } |\mathcal{L}_{skel}| = n - 1.
\end{equation}
Since a tree topology is inherently acyclic, any one-to-one assignment of relations from $\mathcal{L}_{skel}$ to the edges in $E_{tree}$ is guaranteed to be path-consistent. Consequently, we assign the relations in $\mathcal{L}_{skel}$ directly to the skeleton edges, yielding a sparse but connected initial graph $\mathcal{G}_0 = (V, E_{tree})$.

Following the skeleton initialization, we proceed to \textbf{Relation Augmentation}. At this stage, the graph contains only $n-1$ edges; the target complexity requires embedding the remaining relations $\mathcal{L}_{rem}$ into the network. However, arbitrarily placing these constraints can lead to structural biases, where specific nodes accumulate disproportionate constraints, resulting in an uneven distribution of reasoning difficulty.

To mitigate this, we dynamically prioritize unconnected pairs that link sparsely populated regions of the graph. Formally, for any candidate pair $(u, v)$ in the set of unconnected pairs $P_{free}$, we define a selection cost $k(u, v)$ based on the current nodal degrees:
\begin{equation}
k(u, v) = \deg(u) + \deg(v).
\end{equation}
We first sort $P_{free}$ in ascending order of $k$. Then, for each pending relation $r \in \mathcal{L}_{rem}$, we traverse the sorted candidate pairs to find the first $(u, v)$ that satisfies path consistency:
\begin{equation}
\mathrm{IsConsistent}(\mathcal{G} \cup \{(u, v, r)\}) = \textbf{true}.
\end{equation}
If the check passes, the edge is committed, and $(u,v)$ is removed from the pool. This strategy promotes a uniform distribution of complexity by prioritizing structurally simpler connections, thereby maximizing the likelihood that all required relation types are successfully embedded.

The final stage is \textbf{Isomorphism Elimination}. To ensure the benchmark evaluates diverse reasoning patterns rather than memorization of repetitive structures, we must filter out topologically identical graphs. First, we execute a global path consistency check to confirm that the accumulated local constraints result in a globally valid network. Subsequently, we compute a canonical signature $\mathcal{S}$ that identifies the graph's topological structure. To guarantee dataset diversity, we compare $\mathcal{S}$ against a history set $\Sigma$ and retain the graph only if its signature is novel (i.e., $\mathcal{S} \notin \Sigma$).

\zsd{If Algorithm~\ref{alg:graph_generation} returns Failure during any phase, the framework discards only the current randomized attempt rather than the underlying candidate configuration. An outer loop continues to initiate independent trials until the target number of valid graphs is constructed. Naturally, higher difficulty levels impose stricter constraints, necessitating more attempts. For example, successfully generating 40 valid graphs at Difficulty 80 required approximately 800 attempts. Because constraint graph generation is a one-time, parallelizable offline preprocessing step, these discarded attempts incur minimal computational overhead and do not impact the practical usability of the benchmark.}


\subsection{Temporal Reasoning Task Construction}
\label{subsec:construction}

\subsubsection{Extraction of Facts and Questions}
\label{subsubsec:fact_question_extraction}
Based on the temporal constraint graph $\mathcal{G}=(V, E)$ generated in Section~\ref{subsubsec:graph_generation}, \tool{} systematically constructs a dataset of reasoning tasks. For each graph, we formalize the output as a structure $\mathcal{T} = (\mathcal{F}, \mathcal{Q}, \mathcal{A})$, consisting of a shared set of premise facts $\mathcal{F}$, a set of candidate natural language queries $\mathcal{Q}$, and their corresponding ground truth answers $\mathcal{A}$.

The construction process begins with \textbf{Fact Extraction}, which establishes the explicit context $\mathcal{F}$. The edges $E$ in the constraint graph represent the known temporal premises. For every explicit constraint $(u, v, r) \in E$, where $r \in \mathcal{R}_{Allen}$ is the symbolic relation (e.g., \texttt{O}), we map it to its corresponding natural language descriptor $r^{nl}$ (e.g., "overlaps") using a predefined lexicon. The collection of these textual statements constitutes the fact set $\mathcal{F}$, providing the necessary and sufficient conditions to solve the graph.

To formulate meaningful questions that require deductive reasoning rather than simple retrieval, we employ Constraint Propagation. We execute the path consistency algorithm on $\mathcal{G}$ to compute the transitive closure of the network. Let $\mathcal{R}_{inferred}(u, v)$ denote the set of permissible relations between any two events $u$ and $v$ after propagation. We iterate through all event pairs and filter out those already explicitly defined in $E$. The remaining pairs represent implicit relationships that can only be determined by inferring through the chain of facts in $\mathcal{F}$.

From these implicit pairs, we select candidates for \textbf{Question Formulation} to populate $\mathcal{Q}$ and $\mathcal{A}$. To ensure precise evaluation, we focus strictly on deterministic inferences, which are cases where the interaction of constraints narrows the relationship down to a single possibility. A pair $(u, v)$ is selected if and only if $|\mathcal{R}_{inferred}(u, v)| = 1$. Let $r_{true}$ be the unique inferred relation. For each standard Allen relation $r_k \in \mathcal{R}_{Allen}$, we construct a binary verification question $q_k$: ``Is $u$ $r_k^{nl}$ $v$?''. The corresponding ground truth label $a_k$ is determined as:
\begin{equation}
a_k =
\begin{cases}
\text{YES} & \text{if } r_k = r_{true} \\
\text{NO} & \text{otherwise}
\end{cases}.
\end{equation}
The collection of all generated $q_k$ constitutes the query set $\mathcal{Q}$, and the corresponding $a_k$ form the answer set $\mathcal{A}$. This mechanism generates a diverse pool of candidate queries for each inferred fact. 

\subsubsection{Question Templatization}
\label{subsubsec:question_templatization}

To transform the logical task $\mathcal{T}$ into an executable input for Large Language Models, we employ a standardized prompt template. This design not only provides the model with the necessary context, including formal definitions of the 13 Allen relations, but also strictly enforces a structured output format. By requiring the model to generate a JSON object containing a step-by-step \texttt{"reasoning"} trace alongside the \texttt{"final\_answer"}, we enable the parser to extract and verify the underlying logic in the subsequent stage.

Specifically, the prompt is structured into six components. The \texttt{\# Constraint} section defines the reasoner's role and provides context along with restrictions, while \texttt{\# Allen relations} provides formal interval definitions to ensure semantic precision. The \texttt{\# Facts} and \texttt{\# Question} sections present the explicit premises and the target inference query, respectively. Subsequently, \texttt{\# Instructions} enforces the strict JSON schema for the reasoning trace, and \texttt{\# Example} provides a concrete demonstration of the expected output format to ensure parsing compatibility. \zsd{An abbreviated example of the prompt structure is illustrated below, where \texttt{[FACTS]} is instantiated with statements such as ``A after C, B during C,'' and \texttt{[QUESTION]} is instantiated with questions such as ``Is A after B?'' The complete prompt templates and comprehensive task examples are publicly available on our project website~\cite{trace_site}.}

\begin{tcolorbox}[breakable, colback=gray!10, colframe=gray!50, title=\zsd{\textbf{Prompt Template}}, boxrule=0.5pt, left=2pt, right=2pt, top=2pt, bottom=2pt]
\scriptsize \ttfamily \raggedright 
\# Constraint\\
You are a temporal reasoner. Use ONLY event IDs (A, B, C, ...) and these 13 Allen relations: \{before, after, meets, met-by, ...\}.\\
Do not invent events/relations beyond what is stated or logically entailed.\vspace{0.5em}

\# Allen relations\\
Let an event X be an interval [Xs, Xe] with Xs $<$ Xe.\\ 
...\\
- equals: A equals B $<->$ As = Bs and Ae = Be.\vspace{0.3em}

\# Facts: {[}FACTS{]}\vspace{0.3em}

\# Question: {[}QUESTION{]}\vspace{0.3em}

\# Instructions:\\
1) Produce a step-by-step reasoning as a JSON array named "reasoning". Each step is an object \{ "lhs": "X", "rel": "allen\_relation", "rhs": "Y" \} ...\vspace{0.3em}

\# Example\\
\{ "reasoning": [ ... ], "final\_answer": "YES" \}
\end{tcolorbox}


% ...



% ...


\subsection{Trace-Based Verification Oracle}
\label{subsec:oracle}
To rigorously assess whether the model derives the correct answer through valid temporal logic rather than statistical shortcuts, we propose a trace-based verification mechanism. This module executes the reasoning tasks, parses the structured traces, and validates the logical soundness of each derivation step against the ground truth constraints.

The evaluation process begins by feeding the generated prompts into the target LRM. Under the constraints imposed by the prompt template, the LRM produces an internal thinking process (e.g., within <think>...</think>) and then outputs a structured JSON object that contains a step-by-step reasoning chain and the final answer. To handle potential formatting irregularities common in LLM outputs, we employ a robust regex-based parser to extract the reasoning trace $T = [s_1, s_2, \dots, s_k]$ from the JSON object. Each step $s_i$ is formalized as a triplet $(u_i, r_i, v_i)$, representing a claim that "Event $u_i$ has relation $r_i$ to Event $v_i$." Simultaneously, the predicted final answer is extracted and normalized into a standard label $\hat{y} \in \{\text{YES}, \text{NO}, \text{Unsure}\}$.

Unlike traditional methods that typically compare model outputs against a predefined reference or rule-based solution, \tool{} verifies each reasoning step against the algebraic closure implied by the ground-truth constraint network. For every extracted triplet $s_i=(u_i, r_i, v_i)$ from the reasoning chain, we invoke the underlying path-consistency solver on the ground truth graph $\mathcal{G}$ to compute the propagated relation set $\mathcal{R}_{valid}(u_i, v_i)$, which contains all Allen relations that remain feasible between $u_i$ and $v_i$ after constraint propagation from the premise facts. To align verification with the semantics of Allen's algebra, we interpret the model's predicted relation $r_i$ as a relation set $\widehat{\mathcal{R}}_i$: if $r_i$ is a single basic Allen relation, then $\widehat{\mathcal{R}}_i=\{r_i\}$; if the model outputs a disjunctive form such as \texttt{B|M}, we parse it as the corresponding union of basic relations. A step $s_i$ is considered valid if and only if the model's claimed relation set matches the solver-implied closure exactly: 
\begin{equation} 
\widehat{\mathcal{R}}_i \;=\; \mathcal{R}_{valid}(u_i, v_i). 
\end{equation}
This strict criterion filters out ambiguous or incorrect intermediate claims, while avoiding any dependence on a single canonical proof. In other words, any derivation route is accepted as long as every step agrees with the relation set implied by the constraint-solver closure.

Based on these verification results, we classify the model's performance into four distinct categories to provide a comprehensive faithfulness assessment. We define the correctness of the reasoning chain, $C_{reas}$, as a strict conjunction where $C_{reas}=1$ if and only if all steps in $T$ are valid. Combining this with the binary correctness of the final answer ($C_{ans}$), we define the following metrics:
\begin{itemize}
\item \textbf{True Reasoning ($C_{ans}=1, C_{reas}=1$):} The model correctly answers the question based on a logically valid derivation chain.
\item \textbf{Spurious Guessing ($C_{ans}=1, C_{reas}=0$):} The model predicts the correct label, but the reasoning process contains at least one logical error. 
\item \textbf{Answer Misalignment ($C_{ans}=0, C_{reas}=1$):} The model performs sound logical derivation throughout the reasoning chain but produces an incorrect final answer label.
\item \textbf{Complete Failure ($C_{ans}=0, C_{reas}=0$):} The model fails in both the reasoning process and the final prediction.
\end{itemize}
\section{Evaluation} 
In this section, we evaluate the performance of \tool{}. We first outline the experimental setup, including the evaluated LRMs and the dataset generation configuration. Specifically, our evaluation addresses the following research questions:

\begin{itemize}
\item{\textbf{RQ1:} How effective is \tool{} in generating temporal reasoning tasks with controllable difficulty?} 
\item{\textbf{RQ2:} How do different LLM architectures and sizes perform on the temporal reasoning benchmark \benchmark{}?} 
\item{\textbf{RQ3:} To what extent do traditional outcome-based metrics overestimate the reasoning faithfulness of LLMs?} 
\item{\textbf{RQ4:} What are the characteristic failure modes of LLMs in complex temporal reasoning tasks?} 
\end{itemize}

\subsection{Evaluation Setup}
\label{sec:setup}

\subsubsection{Evaluated Models}
We select a representative set of Large Reasoning Models for evaluation, covering both open-weights distilled models and advanced models. Specifically, we include:
\begin{itemize}
    \item \textbf{Open-Weights Distilled Models:} We evaluate the \textbf{DeepSeek-R1-Distill} family to analyze the impact of model scale and base architecture. This includes the Qwen-based variants (\textbf{7B}, \textbf{14B}, \textbf{32B}) and the Llama-based variant (\textbf{Llama-8B}). All distilled models are deployed locally.
    \zsd{\item \textbf{Advanced Models:} We evaluate \textbf{Gemini-2.5-Flash}, \textbf{DeepSeek-R1}, \textbf{GPT-5-mini}, and \textbf{Claude-Sonnet-4.6} via their respective APIs. These models serve as the topline performance baseline.}
\end{itemize}
\zsd{All models are configured with a maximum generation length of 8,192 tokens. To minimize answer stochasticity, the temperature is set to 0 for all models that support this parameter (for some API-based models that do not support temperature settings, we use their default configuration).}

\subsubsection{Dataset Configuration}
We utilize \tool{} to construct \benchmark{}, a temporal reasoning benchmark spanning multiple levels of complexity. We generate reasoning tasks across six distinct difficulty levels: $\mathcal{D}_{\mathrm{tar}} \in \{10, 45, 80, 115, 150, 185\}$. This progression ranges from elementary transitive inference (Difficulty 10) to highly complex temporal networks (Difficulty 185). For each difficulty tier, we generate 40 distinct constraint graphs and sample 200 reasoning questions in total (across these graphs), resulting in 1,200 test samples. To ensure structural consistency across difficulties, we fix the constraint density $\rho = 0.5$ and enforce a strict difficulty alignment with a tolerance threshold $\tau = 0.1$.

\subsection{RQ1: Analysis of Difficulty Controllability}
\label{subsec:rq1}

\begin{table}[htbp]
\centering
\caption{Average Nodes, Edges, and Achieved Difficulty of Constraint Graphs for Target Difficulty from 10 to 185}
\label{tab:stats}
\resizebox{0.75\columnwidth}{!}{
\begin{tabular}{ccccccc}
\toprule
\multirow{2}{*}{\textbf{Metric}} 
& \multicolumn{6}{c}{\textbf{Target Difficulty}} \\
\cmidrule(lr){2-7}
& \textbf{D-10} & \textbf{D-45} & \textbf{D-80} & \textbf{D-115} & \textbf{D-150} & \textbf{D-185} \\
\midrule
\textbf{Nodes}               & 3.00  & 7.00  & 10.00 & 12.00 & 14.00 & 16.00 \\
\textbf{Edges}               & 2.00  & 10.18 & 22.03 & 32.98 & 45.98 & 59.98 \\
\textbf{Achieved Difficulty} & 9.36  & 44.97 & 81.14 & 114.33& 148.32& 185.40 \\
\bottomrule
\end{tabular}
}
\end{table}

\begin{figure}[htbp]
    \centering
    \vspace{-10pt}
    \includegraphics[width=\textwidth]{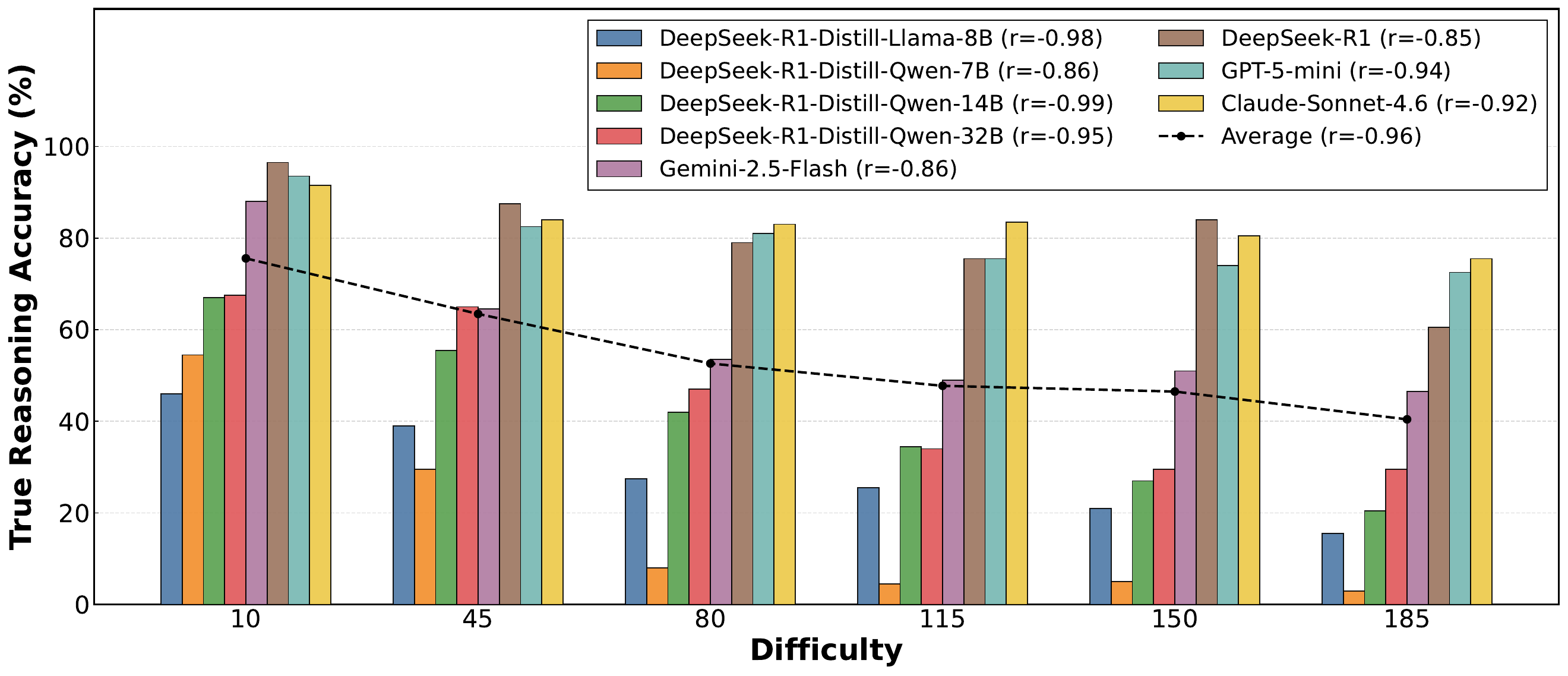}
    \caption{\zsd{True Reasoning Accuracy of Evaluated LRMs Across Six Target Difficulty Levels. The bars represent individual model performance, while the dashed black line indicates the average accuracy trajectory. Pearson correlation coefficients ($r$) are reported in the legend to quantify the negative correlation between difficulty and True Reasoning Accuracy.}}
    \vspace{-10pt}
    \label{fig:RQ1_2}
\end{figure}

To verify whether \tool{} can effectively generate temporal reasoning tasks with controllable difficulty, we conduct a two-step validation. First, we calculate the structural statistics of the generated constraint graphs to verify their alignment with our difficulty specifications. \zsd{Second, we evaluate the performance of eight different LRMs on these datasets across varying difficulty levels to see if the theoretical difficulty score effectively reflects the relative reasoning load on the models, as indicated by actual performance drops.}

\textbf{Structural Analysis of Constraint Graphs.}
Table~\ref{tab:stats} presents the statistics of the generated graphs. As the target difficulty increases from 10 to 185, the graph size increases in line with the configuration. The average number of nodes grows from 3.00 to 16.00, and the number of edges increases from 2.00 to 59.98. The actual achieved difficulty scores are almost identical to the target scores. For example, at the target difficulty of 45, the achieved difficulty is 44.97, and at target difficulty 185, it is 185.40. This confirms that the generation algorithm precisely follows the target difficulty settings.

\textbf{Verification of Difficulty-Performance Correlation.}
Figure~\ref{fig:RQ1_2} illustrates the True Reasoning Accuracy of different models across these difficulty levels. We observe a consistent downward trend for all models as the difficulty increases. For instance, the accuracy of DeepSeek-R1-Distill-Qwen-32B drops from 67.50\% at difficulty 10 to 29.50\% at difficulty 185. Similarly, the smaller model DeepSeek-R1-Distill-Llama-8B starts at 46.00\% and drops to 15.50\% at the hardest level. Even the high-performing GPT-5-mini shows a clear decline, dropping from 93.50\% to 72.50\%. The Pearson correlation coefficients further quantify this trend, ranging from -0.85 to -0.99 across all models, with an average of -0.96. \zsd{This strong negative correlation demonstrates that our metric effectively serves as a relative indicator of the actual reasoning load imposed on the models.}

\begin{tcolorbox}[size=title]{
\textbf{Answer to RQ1:} \tool{} generates temporal reasoning tasks with precise difficulty control. The structural statistics of the generated graphs closely correspond to the target difficulty settings. \zsd{Furthermore, the performance of all evaluated LRMs shows a strong negative correlation ($r_{avg}=-0.96$) with the relative difficulty metric, demonstrating that the generated tasks effectively provide a controllable gradient of difficulty.}}
\end{tcolorbox}

\subsection{RQ2: Benchmarking LLM Performance}
\label{subsec:rq2}
\begin{figure}[htbp]
    \centering
    \includegraphics[width=0.8\textwidth]{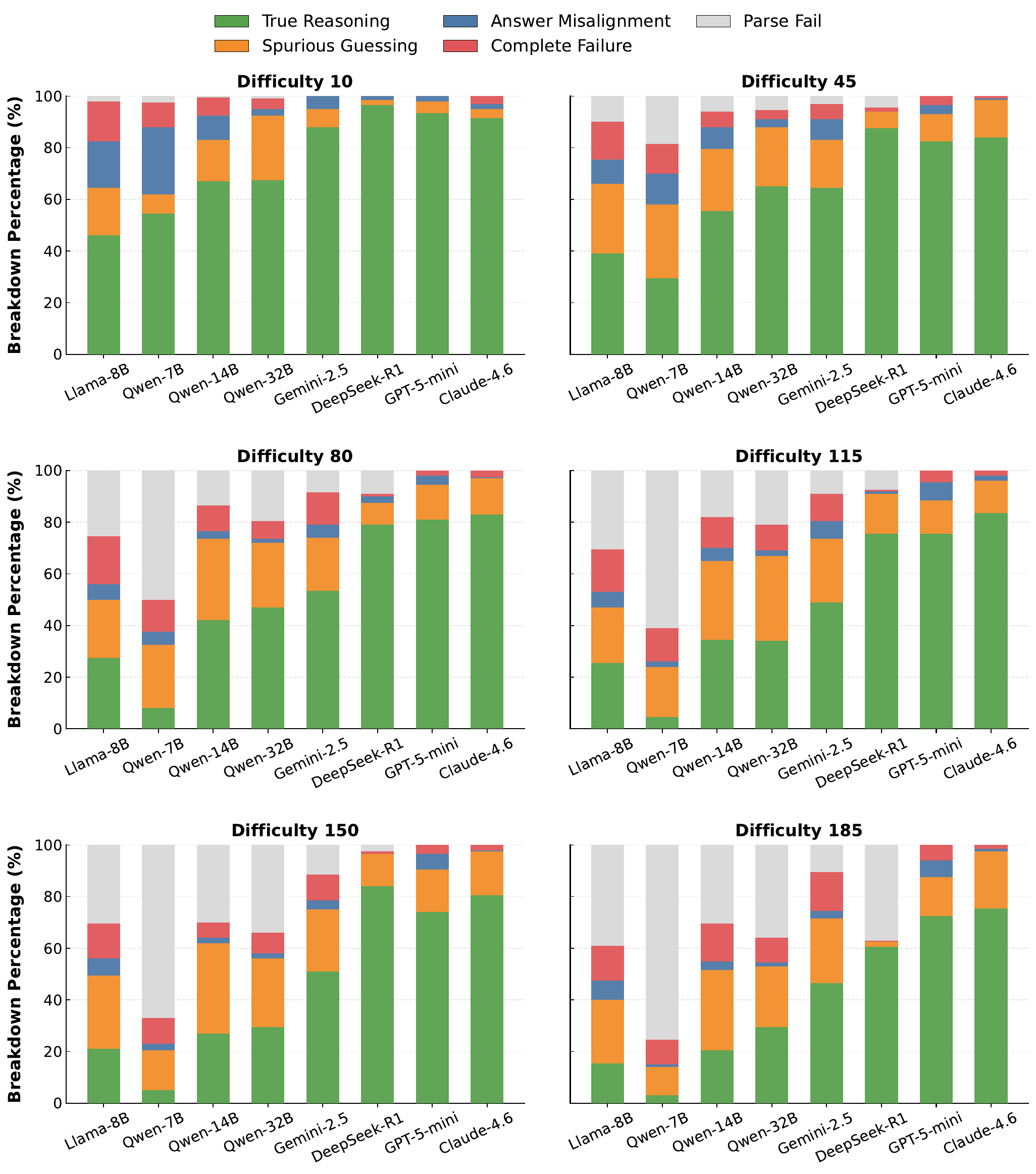}
    \vspace{-10pt}
    \caption{\zsd{Fine-Grained Performance Breakdown of Evaluated LRMs Across Six Difficulty Levels. The 100\% stacked bar charts visualize the composition of model outputs, categorized into True Reasoning, Spurious Guessing, Answer Misalignment, Complete Failure, and Parse Fail.}}
    \label{fig:RQ2_1}
    \vspace{-0.5cm}
\end{figure}

To address RQ2, we evaluate the performance of eight LRMs in \benchmark{}. The goal is to benchmark their temporal reasoning capabilities and understand the impact of different model architectures and parameter scales on temporal reasoning performance. Figure~\ref{fig:RQ2_1} presents a fine-grained breakdown of model outputs, with the green bar (True Reasoning) serving as the primary indicator of effective performance.

\textbf{Impact of Parameter Scaling.}
First, by examining the DeepSeek-R1-Distill-Qwen series, we observe a clear positive correlation between temporal reasoning capability and model scale. At the lowest difficulty (Difficulty 10), the performance gap among the three models is relatively small. The 7B, 14B, and 32B models achieve True Reasoning accuracies of 54.50\%, 67.00\%, and 67.50\%, respectively. However, as the difficulty increases, the 7B model's performance degrades sharply. In contrast, the 14B and 32B models demonstrate much stronger resilience. Although their accuracy also declines, the gap between the two larger models remains narrow, with both significantly outperforming the 7B variant. Overall, the 32B model demonstrates the best robustness, achieving an average accuracy of 45.42\% across all six difficulty levels, followed by the 14B model at 41.08\%, and the 7B model at 17.42\%.

\textbf{Impact of Model Architecture.} 
Second, we examine the influence of base architecture by comparing DeepSeek-R1-Distill-Llama-8B with the DeepSeek-R1-Distill-Qwen series. At the lowest difficulty (Difficulty 10), the Qwen-7B model holds a performance advantage, achieving 54.50\% True Reasoning Accuracy compared to 46.00\% for Llama-8B. However, as the task complexity escalates, Llama-8B demonstrates significantly stronger robustness. Unlike Qwen-7B, which suffers a steep decline, Llama-8B maintains a more stable degradation curve. As difficulty increases, the performance of Llama-8B progressively converges with that of the much larger DeepSeek-R1-Distill-Qwen-14B. The accuracy gap between them narrows consistently from 21.00\% at Difficulty 10 to 16.50\%, 14.50\%, 9.00\%, and 6.00\% across intermediate levels, ultimately shrinking to just 5.00\% at Difficulty 185.

\begin{tcolorbox}[size=title]
\textbf{Answer to RQ2-1:} Reasoning performance generally improves with model scale. However, the base architecture is a critical factor: a smaller model (e.g., Llama-8B) can achieve performance that closely approaches that of larger models (e.g., Qwen-14B) as task difficulty increases.
\end{tcolorbox}

\textbf{Evaluation of Model Capability and Boundaries.}
Finally, we integrate the analysis of response categories with performance boundaries to evaluate the capability thresholds of different models.

First, regarding small-scale models, specifically DeepSeek-R1-Distill-Llama-8B and DeepSeek-R1-Distill-Qwen-7B, we observe a distinct "Answer Misalignment" phenomenon. Across all six difficulty levels, these two models exhibit the highest rate of misalignment, averaging 17.83 and 16.17 samples per level, respectively. This indicates that while they may occasionally derive correct reasoning, they struggle to map this logic to the final label consistently. In terms of performance boundaries, these models hit their limits early. DeepSeek-R1-Distill-Qwen-7B drops rapidly, with its True Reasoning Accuracy dropping to just 8.00\% at Difficulty 80. While DeepSeek-R1-Distill-Llama-8B is relatively more robust, it also struggles at the extreme, achieving only 15.50\% True Reasoning Accuracy at Difficulty 185.

Second, for mid-sized models like DeepSeek-R1-Distill-Qwen-14B and DeepSeek-R1-Distill-Qwen-32B, the primary issue shifts to "Spurious Guessing." These models exhibit a pronounced tendency to predict correct labels without valid reasoning, averaging 56 and 52 spurious samples per level, respectively~(the highest among all models). This behavior leads to a significant inflation of performance metrics when relying solely on answer correctness, a discrepancy we quantify in Section~\ref{subsec:rq3}. Regarding performance boundaries, these models demonstrate extended capability but eventually succumb to complexity. At the maximum difficulty of 185, the True Reasoning Accuracy for the 14B and 32B models falls to 20.50\% and 29.50\%, respectively, indicating a clear ceiling in their ability to handle complex temporal dependencies.

\zsd{Finally, the advanced models establish the performance ceiling for this benchmark, though they exhibit varying operational boundaries. Claude-Sonnet-4.6, DeepSeek-R1, and GPT-5-mini demonstrate exceptional capabilities with average True Reasoning Accuracies of 83.00\%, 80.50\%, and 79.83\%, respectively. Under extreme conditions, their stability diverges. Both Claude-Sonnet-4.6 and GPT-5-mini exhibit strong architectural stability. At Difficulty 185, they maintain high accuracies of 75.50\% and 72.50\%, and neither model produces any Parse Failures throughout the benchmark, indicating robust instruction compliance. In contrast, while DeepSeek-R1 achieves a highly competitive overall average, it encounters a sharper performance drop at the maximum difficulty, falling to 60.50\%. This degradation is primarily driven by context window bottlenecks, where excessive reasoning chain lengths lead to Parse Failures, a specific failure mechanism that we dissect in Section~\ref{subsec:rq4}. Gemini-2.5-Flash performs slightly below this top tier, with an average accuracy of 58.75\%. Moreover, its performance degrades rapidly as task difficulty increases, placing it between mid-sized open-weights models and the highest-performing advanced models.}

\begin{tcolorbox}[size=title] 
\textbf{Answer to RQ2-2:} We identify distinct failure modes across model sizes. Small models exhibit higher levels of Answer Misalignment, while mid-sized models are prone to Spurious Guessing. Advanced models remain robust but encounter difficulties at extreme complexity, largely due to the challenge of processing extended reasoning contexts.
\end{tcolorbox}


\subsection{RQ3: Assessment of Reasoning Faithfulness}
\label{subsec:rq3}
To address RQ3, we investigate the prevalence of Spurious Guessing, which refers to instances where models predict the correct final label despite failing to provide a valid reasoning trace. This analysis aims to quantify the discrepancy between outcome-based accuracy and true reasoning performance. If a model relies on statistical shortcuts rather than logical deduction, outcome-based metrics serve as an inflated proxy for actual performance. Table~\ref{tab:RQ3_1} illustrates the Spurious Guessing Rate across different difficulty levels. We observe three distinct behavioral patterns:

\begin{table}[htbp]
\centering
\caption{\zsd{Spurious Guessing Rate Across Six Difficulty Levels. Llama-8B and Qwen-7B/14B/32B refer to DeepSeek-R1-Distill models.}}
\label{tab:RQ3_1}
\resizebox{0.75\columnwidth}{!}{
\begin{tabular}{lcccccc}
\toprule
\multirow{2}{*}{\textbf{Model}} 
& \multicolumn{6}{c}{\textbf{Target Difficulty}} \\
\cmidrule(lr){2-7}
& \textbf{D-10} & \textbf{D-45} & \textbf{D-80} & \textbf{D-115} & \textbf{D-150} & \textbf{D-185} \\
\midrule
\textbf{Llama-8B}   & 18.50\% & 27.00\% & 22.50\% & 21.50\% & \textbf{28.50\%} & 24.50\% \\
\rowcolor{gray!20} 
\textbf{Qwen-7B}    & 7.50\%  & \textbf{28.50\%} & 24.50\% & 19.50\% & 15.50\% & 11.00\% \\
\textbf{Qwen-14B}   & 16.00\% & 24.00\% & 31.50\% & 30.50\% & \textbf{35.00\%} & 31.00\% \\
\rowcolor{gray!20} 
\textbf{Qwen-32B}   & 25.00\% & 23.00\% & 25.00\% & \textbf{33.00\%} & 26.50\% & 23.50\% \\
\textbf{Gemini-2.5-Flash} & 7.00\%  & 18.50\% & 20.50\% & 24.50\% & 24.00\% & \textbf{25.00\%} \\
\rowcolor{gray!20} 
\textbf{DeepSeek-R1} & 2.00\%  & 6.50\%  & 8.50\%  & \textbf{15.50\%} & 12.50\% & 2.00\%  \\
\textbf{GPT-5-mini} & 4.50\%  & 10.50\% & 13.50\% & 13.00\% & \textbf{16.50\%} & 15.00\% \\
\rowcolor{gray!20} 
\textbf{Claude-Sonnet-4.6} & 3.50\%  & 14.50\%  & 14.00\%  & 12.50\% & 17.00\% & \textbf{22.00\%} \\
\bottomrule
\end{tabular}
}
\end{table}

\textbf{High Overestimation in Mid-Sized Models.} 
The most significant inflation of performance metrics appears in mid-sized distilled models. DeepSeek-R1-Distill-Qwen-14B and 32B exhibit consistently high spurious guessing rates, averaging 28.00\% and 26.00\% respectively across all difficulties. Notably, the 14B model's spurious rate climbs with complexity, peaking at 35.00\% at Difficulty 150. This indicates that these models possess sufficient parameter capacity to exploit statistical shortcuts for label prediction, yet they frequently fail to construct sound and complete reasoning traces. Consequently, relying solely on answer correctness for this model class yields the most misleading assessment.

\textbf{Instability in Small Models.}
The smaller DeepSeek-R1-Distill-Qwen-7B displays a volatile "early-peak" pattern. Its spurious guessing surges to 28.50\% at Difficulty 45 but declines sharply as complexity increases. This decline does not reflect improved faithfulness; rather, it aligns with the performance collapse observed in RQ2. As the model becomes unable to predict correct labels reliably under high complexity, the occurrence of spurious successes naturally diminishes.

\textbf{High Faithfulness in Advanced Models.}
In contrast, DeepSeek-R1 demonstrates superior reasoning faithfulness. Its spurious guessing rate remains low, with an average of 7.83\%, showing only minor fluctuations in the intermediate difficulty range. This confirms that advanced models are highly consistent: when they output a correct answer, it is almost invariably grounded in a valid logical derivation. \zsd{Similarly, GPT-5-mini and Claude-Sonnet-4.6 maintain relatively low and stable spurious rates (averaging 12.17\% and 13.92\%, respectively), further supporting that stronger reasoners depend less on guesswork. Gemini-2.5-Flash also shows a moderate spurious rate of 19.92\%, remaining more faithful than the mid-sized distilled models.}

\begin{tcolorbox}[size=title]
\textbf{Answer to RQ3:} Traditional outcome-based metrics significantly overestimate model performance, particularly for mid-sized models which show the highest average spurious guessing rates (up to 28.00\%). In contrast, advanced models demonstrate high consistency between answer correctness and logical validity, highlighting the necessity of \tool{}'s trace-based verification for precise evaluation.
\end{tcolorbox}

\subsection{RQ4: Diagnosis of Failure Modes}
\label{subsec:rq4}

To address RQ4, we investigate the characteristic failure modes of LLMs in complex temporal reasoning tasks. We conduct a manual inspection of a representative subset of failure cases sampled from the experimental results. Based on this analysis, we summarize the observed errors into two primary categories: \textit{Logical Failures}, where models generate valid formats but yield incorrect reasoning or answers, and \textit{Structural Failures}, where models fail to produce parsable outputs due to format violations or context limitations.

\subsubsection{Logical Failures}
\label{subsubsec:logical_failures}

Through a detailed inspection of samples where models produce validly formatted outputs but yield incorrect reasoning or answers, we categorize the observed logical failures into three primary classes: \textbf{Direct Inference Error}, \textbf{Reasoning Stagnation}, and \textbf{Hallucination under Complexity}.

The first two types represent deficiencies in reasoning capability and efficiency. The most common error across all models is \textbf{Direct Inference Error}. In this case, the model draws an incorrect conclusion from the given premises by misapplying transitivity or temporal constraints, which leads to an incorrect final answer. The second type is \textbf{Reasoning Stagnation}. In these cases, the model generates a lengthy reasoning trace that appears structurally valid but merely rephrases or permutes the provided facts. This failure to progress logically prevents the model from deriving the target conclusion, effectively forcing it to guess the final answer despite a long chain of thought.

The third type is \textbf{Hallucination under Complexity}, which manifests as a breakdown in the model's grounding to the provided context. We observe that models occasionally introduce undefined elements into the JSON reasoning trace, including non-existent events (e.g., "As", "?") or non-standard relations that are not defined in the prompt (e.g., "<", "=", "less\_than", "cannot overlap"). This phenomenon appears across all model scales, suggesting that under high complexity, models may hallucinate external knowledge or symbols to bridge logical gaps. 

\begin{table}[htbp]
\centering
\caption{\zsd{Distribution of Logical Failure Modes Across Evaluated LRMs.}}
\label{tab:logical_failures}
\resizebox{\columnwidth}{!}{
\begin{tabular}{lccc}
\toprule
\textbf{Model} & \textbf{Direct Inference Error} & \textbf{Reasoning Stagnation} & \textbf{Hallucination} \\
\midrule
\textbf{Llama-8B}          & 68.92\% & 15.97\% & 15.10\% \\
\rowcolor{gray!20}
\textbf{Qwen-7B}           & 50.00\% & 17.65\% & 32.35\% \\
\textbf{Qwen-14B}          & 81.57\% & 9.41\%  & 9.02\%  \\
\rowcolor{gray!20}
\textbf{Qwen-32B}          & 94.77\% & 4.04\%  & 1.19\%  \\
\textbf{Gemini-2.5-Flash}  & 96.83\% & 2.68\%  & 0.49\%  \\
\rowcolor{gray!20}
\textbf{DeepSeek-R1}       & 79.65\% & 3.54\%  & 16.81\% \\
\textbf{GPT-5-mini}        & 89.26\% & 9.09\%  & 1.65\%  \\
\rowcolor{gray!20}
\textbf{Claude-Sonnet-4.6} & 96.08\% & 3.92\%  & 0.00\%  \\
\bottomrule
\end{tabular}
}
\end{table}

\zsd{To further substantiate these observations, we developed automated scripts based on our initial manual inspection to classify all Logical Failures. As detailed in Table~\ref{tab:logical_failures}, the quantitative distribution confirms our qualitative analysis. \textbf{Direct Inference Error} is the overwhelmingly dominant failure mode, accounting for at least 50\% of logical errors across every evaluated model. Meanwhile, \textbf{Reasoning Stagnation} remains a persistent baseline issue across the distilled models. Finally, while \textbf{Hallucination under Complexity} affects all scales, smaller architectures like Qwen-7B struggle with it most significantly (32.35\%).} A representative example of this hallucinatory behavior is illustrated below:

\begin{tcolorbox}[breakable, colback=gray!10, colframe=gray!50, title=\textbf{Case Study: Hallucination in Reasoning Trace (GPT-5-mini, Difficulty = 115)}, boxrule=0.5pt]
\small
\begin{verbatim}
...
{"lhs": "J", "rel": "overlapped-by", "rhs": "K"}, 
\end{verbatim}
\vspace{-0.5em}
\begingroup
\color{red!80!black}
\begin{verbatim}
{"lhs": "Cs", "rel": "less_than", "rhs": "Gs"},
{"lhs": "Cs", "rel": "less_than", "rhs": "Hs"}
\end{verbatim}
\endgroup
\begin{verbatim}
...
\end{verbatim}
\end{tcolorbox}

\begin{tcolorbox}[size=title]
\textbf{Answer to RQ4-1:} Logical failures manifest in three distinct modes: (1) Direct Inference Errors, which are the most prevalent; (2) Reasoning Stagnation, where models parrot facts without logical progress; and (3) Hallucinations, where models invent undefined events or relations.
\end{tcolorbox}

\subsubsection{Structural Failures}
\label{subsubsec:structural_failures}

Through an inspection of samples where models failed to produce parsable outputs, we categorize the observed structural failures into two primary classes: \textbf{Format Non-Compliance} and \textbf{Context Window Exhaustion}. These failures represent a breakdown in the model's instruction-following capabilities and resource management when facing extreme reasoning complexity.

The first type is \textbf{Format Non-Compliance}, which is predominantly observed in small and mid-sized models. In these instances, the model fails to strictly adhere to the required JSON schema. The second type is \textbf{Context Window Exhaustion}, where the generation is truncated due to exceeding the token limit. We identify two distinct mechanisms driving this exhaustion. The first mechanism is \textbf{Reasoning Explosion}, primarily driven by the exponential growth of the necessary inference chain as difficulty rises. Notably, DeepSeek-R1's context failures are primarily attributed to this phenomenon. The incidence spikes dramatically at Difficulty 185, affecting 74 out of 200 samples, whereas other difficulty levels record a maximum of only 18 cases. \zsd{The second mechanism is \textbf{Degenerative Loops}, typically found in small and mid-sized distilled models, as well as Gemini-2.5-Flash.} As task difficulty escalates, these models often fall into a "cognitive collapse," infinitely repeating a single reasoning step or phrase until the context window is exhausted. This phenomenon is similar to observations reported in prior studies~\cite{DBLP:journals/corr/abs-2412-21187, DBLP:journals/corr/abs-2505-23480}. 

\begin{table}[htbp]
\centering
\caption{\zsd{Distribution of Structural Failure Modes Across Evaluated LRMs (Based on sampled parse-failure cases).}}
\label{tab:structural_failures}
\resizebox{\columnwidth}{!}{
\begin{tabular}{lccc}
\toprule
\textbf{Model} & \textbf{Format Non-Compliance} & \textbf{Reasoning Explosion} & \textbf{Degenerative Loops} \\
\midrule
\textbf{Llama-8B}          & 20.00\% & 5.00\%   & 75.00\% \\
\rowcolor{gray!20}
\textbf{Qwen-7B}           & 10.00\% & 15.00\%  & 75.00\% \\
\textbf{Qwen-14B}          & 0.00\%  & 5.00\%   & 95.00\% \\
\rowcolor{gray!20}
\textbf{Qwen-32B}          & 0.00\%  & 10.00\%  & 90.00\% \\
\textbf{Gemini-2.5-Flash}  & 0.00\%  & 0.00\%   & 100.00\% \\
\rowcolor{gray!20}
\textbf{DeepSeek-R1}       & 0.00\%  & 100.00\% & 0.00\%  \\
\bottomrule
\end{tabular}
}
\vspace{-10pt}
\end{table}

\zsd{To quantify these structural breakdowns, we manually inspected 20 randomly sampled parse-failure cases per model (excluding GPT-5-mini and Claude-Sonnet-4.6, which recorded zero parse failures). As summarized in Table~\ref{tab:structural_failures}, format issues are primarily confined to smaller models. \textbf{Degenerative Loops} constitute the most severe problem for open-weights distilled models and Gemini-2.5-Flash, whereas DeepSeek-R1's structural failures are entirely (100.00\%) driven by \textbf{Reasoning Explosion}.} A representative example of such a loop is illustrated below:

\begin{tcolorbox}[breakable, colback=gray!10, colframe=gray!50, title=\textbf{Case Study: Degenerative Loop (DeepSeek-R1-Distill-Qwen-7B, Difficulty = 185)}, boxrule=0.5pt]
\small
\begin{verbatim}
... But J is equal to K, which is from I's start to I's end.
So I's start is before A's end, and I's end is J's end, which is K's end.
So I is from (before A's end) to K's end.
\end{verbatim}
\vspace{-0.5em}
\begingroup
\color{red!80!black}
\begin{verbatim}
But K is equal to J, which is equal to I's finish.
So I's finish is K's end.
So I is from (before A's end) to K's end.

But K is equal to J, which is equal to I's finish.
... 
[Repeated indefinitely until context limit]
\end{verbatim}
\endgroup
\end{tcolorbox}

\begin{tcolorbox}[size=title]
\textbf{Answer to RQ4-2:} Structural failures reveal scale-dependent limitations. \zsd{\textbf{Format Non-Compliance} is mainly concentrated in small models, whereas \textbf{Degenerative Loops} are prevalent in small and mid-sized models as well as Gemini-2.5-Flash}. In contrast, advanced models like DeepSeek-R1 primarily suffer from \textbf{Reasoning Explosion}, driven by valid but excessive deductive chains.
\end{tcolorbox}
\section{Discussion}
\textbf{The Shift to Process-Centric Verification.} This study highlights a fundamental limitation in current benchmarks: the inability to distinguish between genuine deductive reasoning and the "Clever Hans" effect, where models rely on statistical correlations rather than logic. Our findings suggest that as LRMs integrate into high-stakes domains, outcome-based accuracy is an insufficient metric for safety and reliability. Effective evaluation requires treating reasoning as a verifiable trajectory rather than a singular output. \tool{} demonstrates that rigorous, trace-level auditing serves as a pivotal mechanism to ensure models are not merely mimicking logic, but actively executing it.

\zsd{\textbf{Guidance for LRM Development.} Our diagnosed failure modes may inform future LRM training and alignment. First, to address Spurious Guessing, developers can introduce Process Reward Models alongside widely-used Outcome Reward Models to explicitly supervise intermediate deductive steps, rather than relying solely on final answers. Second, to mitigate Answer Misalignment, strict consistency penalties should be applied for mismatches between the generated reasoning trace and the final label. Finally, for Reasoning Explosion, a task-complexity estimation mechanism can be introduced to dynamically allocate and constrain the length of the CoT, thereby effectively reducing redundancy and context exhaustion.}

\zsd{\textbf{Evaluating Pure Reasoning.} \tool{} focuses exclusively on the intrinsic logical capabilities of LRMs. While practical applications often delegate complex constraint satisfaction to external solvers, allowing such tool augmentation in this context would shift the evaluation target from intrinsic deduction to tool-use proficiency. By enforcing a strict, self-contained reasoning environment across varying complexities, \tool{} functions as a targeted diagnostic instrument. It is precisely this isolation that allows us to expose and analyze fundamental cognitive bottlenecks, such as degenerative loops and reasoning explosions, which would otherwise remain masked if the reasoning load were offloaded.}

\textbf{Trade-offs in Controlled Synthesis.} To ensure precise difficulty gradients, we prioritize logical control over naturalistic diversity. By synthesizing tasks from strictly defined algebraic rules, we isolate pure deductive reasoning capacity from the confounding variables of natural language. We acknowledge that this design lacks the semantic ambiguity, linguistic nuance, and "noisy" context inherent in real-world communication. Consequently, \tool{} serves as a diagnostic instrument for intrinsic reasoning robustness, rather than a complete substitute for benchmarks grounded in unstructured, open-domain scenarios.
\section{Conclusion}
In this work, we introduce \tool{}, a framework leveraging Allen's Interval Algebra to generate difficulty-controllable tasks for benchmarking LRMs. Based on this framework, we construct \benchmark{}, a comprehensive graded benchmark designed to systematically probe reasoning boundaries. Our evaluation confirms a precise alignment between task complexity and model performance, while exposing the inadequacy of outcome-based metrics due to the prevalence of \textit{Spurious Guessing} in mid-sized models. Furthermore, we diagnose distinct, scale-dependent failure modes under extreme complexity: \textit{Degenerative Loops} in smaller architectures and \textit{Reasoning Explosion} in advanced models. These findings demonstrate the effectiveness of \tool{} and establish a rigorous foundation for assessing the true reasoning boundaries of AI systems.

\section{Data Availability}
The code, data, and additional information relevant to this study are available at \url{https://anonymous.4open.science/r/TRACE-2061} and our project website~\cite{trace_site}.

{\bibliographystyle{ACM-Reference-Format}
\bibliography{ref}}
\end{document}